\documentclass[english,english,12pt,onecolumn, draftcls]{IEEEtran}
\usepackage[T1]{fontenc}
\usepackage[latin9]{inputenc}
\usepackage{amsthm}
\usepackage{amsmath}
\usepackage{amssymb}
\usepackage{graphicx}
\PassOptionsToPackage{version=3}{mhchem}
\usepackage{mhchem}

\makeatletter
  \theoremstyle{plain}
  \newtheorem{lem}{\protect\lemmaname}
  \theoremstyle{remark}
  \newtheorem{rem}{\protect\remarkname}
  \theoremstyle{plain}
  \newtheorem{prop}{\protect\propositionname}

\ifCLASSINFOpdf
\else
\fi
\makeatother

\usepackage{babel}
\providecommand{\lemmaname}{Lemma}
\providecommand{\propositionname}{Proposition}
\providecommand{\remarkname}{Remark}

\begin{document}

\title{Constrained Codes for Joint Energy and Information Transfer}

\author{Ali Mohammad Fouladgar, Osvaldo Simeone, and Elza Erkip %
\thanks{A.M. Fouladgar and O. Simeone are with the CWCSPR, New Jersey Institute
of Technology, Newark, NJ 07102 USA (e-mail: \{af82,osvaldo.simeone\}@njit.edu).

E. Erkip is with the Department of Electrical Engineering, Polytechnic
Institute of New York University, Brooklyn, NY (email: elza@poly.edu).%
}}
\maketitle
\begin{abstract}
In various wireless systems, such as sensor RFID networks and body
area networks with implantable devices, the transmitted signals are
simultaneously used both for information transmission and for energy
transfer. In order to satisfy the conflicting requirements on information
and energy transfer, this paper proposes the use of constrained run-length
limited (RLL) codes in lieu of conventional unconstrained (i.e., random-like)
capacity-achieving codes. The receiver's energy utilization requirements
are modeled stochastically, and constraints are imposed on the probabilities
of battery underflow and overflow at the receiver. It is demonstrated
that the codewords' structure afforded by the use of constrained codes
enables the transmission strategy to be better adjusted to the receiver's
energy utilization pattern, as compared to classical unstructured
codes. As a result, constrained codes allow a wider range of trade-offs
between the rate of information transmission and the performance of
energy transfer to be achieved.
\end{abstract}

\section{Introduction\label{sec:Introduction}}

Various modern wireless systems, such as sensor RFID networks \cite{Chawla}
and body area networks with implantable devices \cite{ZhangHackworth}-\cite{Nikitin},
challenge the conventional assumption that the energy received from
an information bearing signal cannot be reused. For instance, implantable
devices can be powered by the received radio signal, hence alleviating
the need for a battery and reducing significantly the size of the
devices. This realization has motivated a number of research groups
to investigate the design of wireless systems under joint information
and energy transfer requirements%
\footnote{It is worth noting that wireless energy transfer, has a long history
\cite{Brown} and is available commercially (see, e.g., \cite{y}\cite{z}). %
}.

The research activity in this area has focused so far on optimal resource
allocation in the presence of information and energy transfer for
various network topologies. Specifically, reference \cite{Varshney}
studied a single point-to-point channel, while \cite{Grover}\cite{Varshney1}
investigated power allocation for a set of parallel point-to-point
channels under energy transfer and information rate constraints. The
optimization of beamforming strategies under the same criteria was
studied in \cite{zhang bc}-\cite{Huang} for multiantenna broadcast
channels and for two-user multiantenna interference channels in \cite{Park}.
Optimal resource allocation assuming wireless energy transfer was
also investigated in \cite{Huang-1} for cellular systems, in \cite{Zhang}-\cite{Nasir}
for relay systems, in \cite{Popovski} for two-way interactive channels,
and in \cite{Foulad} for graphical multi-hop networks. Considerations
on the design of the receiver under the constraint that, when harvesting
energy from the antenna, the receiver is not able to use the same
signal for information decoding, can be found in \cite{Liu}. 

Unlike all prior work summarized above, this work focuses on the \textit{code
design} for systems with joint information and energy transfer. We
focus on a point-to-point link as shown in Fig. \ref{fig:fig1}, in
which the receiver's energy requirements are modeled as a random process.
The statistics of this process generally depend on the specific application
to be run at the receiver, e.g., sensing or radio transmission. The
performance in terms of energy transfer is measured by the probabilities
of overflow and underflow of the battery at the receiver. The probability
of overflow measures the efficiency of energy transfer by accounting
for the energy wasted at the receiver. Instead, the probability of
underflow is a measure of the fraction of the time in which the application
run at the receiver is in outage due to the lack of energy.

Classical codes, which are designed with the only aim of maximizing
the information rate, are unstructured (i.e., random-like), see, e.g.,
\cite{Cover}. As a result, they do not allow to control the timing
of the energy transfer, and hence to optimize the probability of overflow
and underflow. With this in mind, here it is proposed to adopt constrained
run-length limited (RLL) codes \cite{Marcus} in lieu of conventional
unconstrained codes. The constraints defining RLL codes ensure that
the code does not includes bursts of energy either too frequently,
thus limiting battery overflow, or too infrequently, thus controlling
battery underflow. Constrained RLL codes have been traditionally studied
for applications related to magnetic and optical storage \cite{Marcus}.
The application to the problem at hand of energy transfer has been
previously studied in the context of point-to-point RFID systems in
\cite{Barbero}, although no analysis of the information-energy trade-off
was provided. In contrast, in this work, a thorough analysis is provided
of the interplay between information rate ad energy transfer in terms
of probabilities of battery overflow and underflow. The analysis reveals
that, by properly choosing the parameters that define RLL codes depending
on the receiver's utilization requirements, constrained codes allow
to greatly improve the system performance in terms of simultaneous
energy and information transfer. 

The reminder of this paper is organized as follows. In Sec. \ref{sec:System-Model},
the system model is introduced along with performance criteria. In
Sec. \ref{sub:Unconstrained-Codes} and Sec. \ref{sub:Constrained-Codes},
we study the performance of classical unconstrained codes and of constrained
RLL codes, respectively, in terms of energy and information transfer.
Sec. \ref{sec:Numerical-Results} presents numerical results. Finally,
some concluding remarks can be found in Sec. \ref{sec:Conclusions}.

\section{System Model\label{sec:System-Model}}

\begin{figure}[h!]
\centering \includegraphics[clip,scale=0.7]{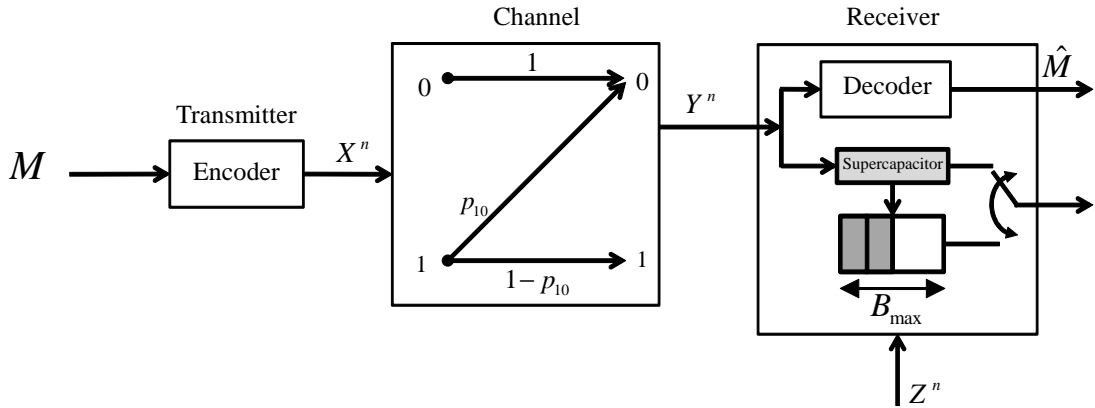}

\caption{Point-to-point link with information and energy transfer.}

\label{fig:fig1}
\end{figure}

We consider the point-to-point channel illustrated in Fig. \ref{fig:fig1}.
We assume that at each discrete time $i$, the transmitter can either
send an \textquotedblleft{}on\textquotedblright{} symbol $X_{i}=1$,
which costs one unit of energy, or an \textquotedblleft{}off\textquotedblright{}
signal $X_{i}=0$, which does not require any energy expenditure.
The receiver either obtains an energy-carrying signal, which is denoted
as $Y_{i}=1$, or receives no useful energy, which is represented
as $Y_{i}=0$. The channel is memoryless, and has transition probabilities
as shown in Fig. \ref{fig:fig1}. Accordingly, $p_{10}$ represents
the probability that energy is lost when propagating between transmitter
and receiver%
\footnote{A more general model would allow also for a non-zero probability $p_{01}$
of receiving energy when no energy is transmitted. This could be interpreted
as the probability of harvesting energy from the environment (see
\cite{Popovski}). We do not consider this extension in this work.%
}. At the receiver side, upon reception of an energy-carrying signal
$Y_{i}=1$, the energy contained in the signal is harvested. The harvested
energy is temporarily held in a supercapacitor and, if not used in
the current time interval $i$, is stored in a battery, whose capacity
limited to $B_{\max}$ energy units (see, e.g., \cite{x}). 

\begin{figure}[h!]
\centering \includegraphics[clip,scale=0.7]{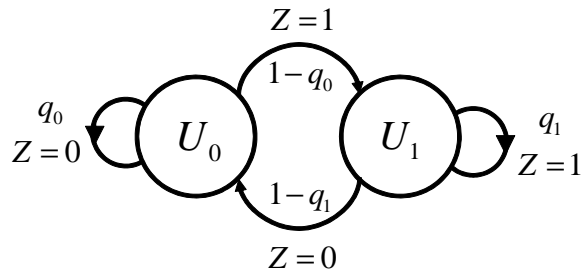}

\caption{Energy utilization model at the receiver.}

\label{fig:fig2}
\end{figure}

The receiver's energy utilization is modeled as a stochastic process
$Z_{i}\in\left\{ 0,1\right\} $, so that $Z_{i}=1$ indicates that
the receiver requires one unit of energy at time $i$, while $Z_{i}=0$
implies that no energy is required by the receiver at time $i$. This
process is not known at the transmitter and evolves according to the
Markov chain shown in Fig. \ref{fig:fig2}. Note that adopting a Markov
model to account for the time variations of energy usage is a standard
practice (see, e.g., \cite{O'Neill} and references therein). Accordingly,
when in state $U_{0}$, there may be bursts of consecutive time instants
in which no energy is required (i.e., $Z_{i}=0$); while, when in
state $U_{1}$, there may be bursts of consecutive time instants in
which energy is required (i.e., $Z_{i}=1$). The probability that
$Z_{i}=0$ when in state $U_{0}$ is referred to as $q_{0}$ and the
probability that $Z_{i}=1$ in state $U_{1}$ is denoted as $q_{1}$.
We observe that the average length of bursts of symbols in which $Z_{i}=j$
in state $U_{j}$ is given by $1/(1-q_{j})$ for $j\in\left\{ 0,1\right\} $.
Also, it is remarked that, when $q_{0}=1-q_{1}$, the energy usage
model becomes a memoryless process with $\Pr[Z_{i}=1]=1-q_{0}=q_{1}$.

Due to the finite capacity of the battery, there may be battery overflows
and underflows. An overflow event takes place when energy is received
and stored in the supercapacitor (i.e., $Y_{i}=1$), but is not used
by the receiver (i.e., $Z_{i}=0$) and the battery is full (i.e.,
$B_{i}=B_{\max}$), so that the energy unit is lost; instead, an underflow
event occurs when energy is required by the receiver (i.e., $Z_{i}=1$)
but the supercapacitor and the battery are empty (i.e., $B_{i}=0$
and $Y_{i}=0$). In the rest of this section we define all the parts
of the system in Fig. \ref{fig:fig1} in detail.

\subsection{Transmitter\label{sub:Transmitter}}

The transmitter aims at communicating a message $M$, uniformly distributed
in the set $\left[1:2^{nR}\right]$, reliably to the decoder, while
at the same time guaranteeing desired probabilities of battery overflow
and underflow (see Sec. \ref{sub:Receiver}). Note that $n$ is the
codeword length and $R$ represents the information rate in bits per
channel use, while the constraints on the probabilities of overflow
and underflow represents the requirements on energy transfer. As discussed
in Sec. \ref{sec:Introduction}, in this work, we investigate the
performance in terms of information and energy transfer achievable
of conventional unconstrained codes and of RLL codes. We introduce
RLL codes next following \cite{Marcus}.

\begin{figure}[h!]
\centering \includegraphics[clip,scale=0.8]{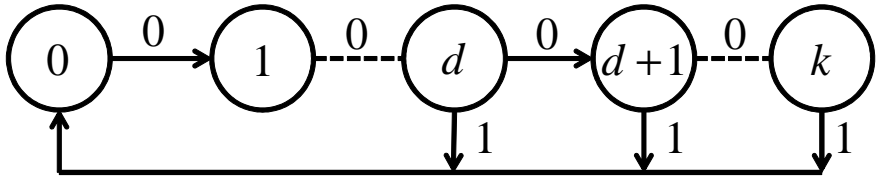}

\caption{The codewords of a type-0 $(d,k)$-RLL code must be outputs of the
shown finite-state machine. A type-1 $(d,k)$-RLL constrained code
is instead obtained by substituting all ``0'' for ``1'' and vice
versa.}

\label{fig:fig3}
\end{figure}

The codewords $x^{n}(m)$, with $m\in\left[1:2^{nR}\right]$, of a
type-$i$ RLL code satisfy run-length constraints on the number of
consecutive symbols $i$, where $i=0$ or $i=1$%
\footnote{Classical RLL codes as discussed in, e.g., \cite{Marcus} are type-0,
but here we find it useful to extend the definition to include also
type-1 RLL codes.%
}. To elaborate, let $d$ and $k$ be integers such that $0\leq d\leq k$.
We say that a finite length binary sequence $x^{n}(m)$ satisfies
the type-0 $(d,k)$-RLL constraint if the following two conditions
hold (see Fig. \ref{fig:fig3}):
\begin{itemize}
\item the runs of 0's have length at most $k$, and
\item the runs of 0's between successive 1's have length at least $d$;
note that the first and last runs of 0's are allowed to have lengths
smaller than $d$.
\end{itemize}
Therefore, a type-0 $(d,k)$-RLL code is such that the codewords include
sufficiently long stretches of zero-energy symbols 0, via the selection
of $d$, thus limiting battery overflow, but not too infrequently,
via $k$, thus partly controlling also battery underflow. As a result,
type-0 $(d,k)$-RLL codes are suitable for overflow-limited regimes
in which controlling overflow events is most critical. An example
of a sequence satisfying the type-0 $(d,k)=(2,7)$-RLL constraint
is $x^{n}(m)=00100001001000000010$ where $n=20$. Overall, the set
of all sequences $x^{n}(m)$ satisfying a type-0 $(d,k)$-RLL constraint
is then described by all the possible $n$-bit outputs of the finite
state machine in Fig. $\ref{fig:fig3}$, where the outputs are shown
by the binary labels of the directed edges. Note that finite-state
machine consists of $k+1$ states (the numbered circles) and the initial
state is arbitrary. 

A type-1 $(d,k)$-RLL code is defined in the same way, upon substitution
of all ``0'' for ``1'' and vice versa in the edge labels of Fig.
\ref{fig:fig3}. Therefore, type-1 $(d,k)$-RLL codes allow one to
control the stretches of ``1'' symbols in the codewords, and are
hence well suited for underflow-limited regimes, in which controlling
the probability of underflow is most important.

\subsection{Receiver\label{sub:Receiver}}

Transmitter and receiver communicate over the binary channel shown
in Fig. \ref{fig:fig1} with the probability of $p_{10}$ of flipping
symbol \textquotedblleft{}1\textquotedblright{} to symbol \textquotedblleft{}0\textquotedblright{}.
As mentioned, this probability can be interpreted in terms of energy
losses across the channel. The received signal $Y^{n}$ is used by
the decoder both to decode the information message $M$ encoded via
the constrained code at the transmitter and to perform energy harvesting.
The harvested energy is used to fulfill the energy requirements of
the receiver as dictated by the process $Z^{n}$, where the requirements
are given in terms of probability of overflow and underflow. This
is discussed next.

Let $B_{i}$ denote the number of energy units available in the battery
at time $i$. At the $i$th time period, the decoder first receives
signal $Y_{i}$, and stores its energy (if $Y_{i}=1$) temporarily
in a supercapacitor (see Fig. \ref{fig:fig1}). Then, if $Z_{i}=1$,
the receiver attempts to draw one energy unit from the supercapacitor
or, if the latter is empty, from the battery. If the energy in the
supercapacitor is not used, it is stored in the battery in the next
time slot. As a result, the amount of energy in the battery evolves
as 
\begin{equation}
B_{i+1}=\min\left(B_{max},\left(B_{i}+Y_{i}-Z_{i}\right)^{+}\right),\label{eq:battery}
\end{equation}
where $\left(a\right)^{+}=\max\left(0,a\right)$. 

\begin{figure}[h!]
\centering \includegraphics[clip,scale=0.5]{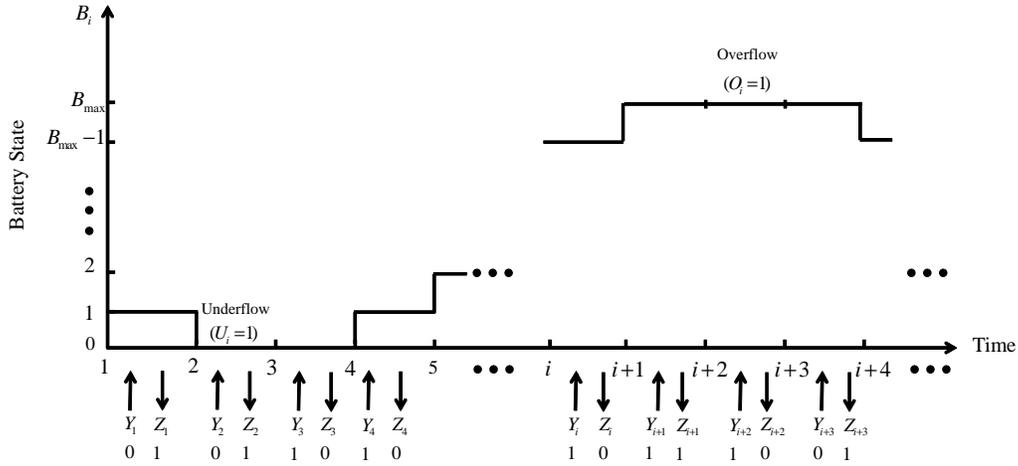} \caption{A sample path of the evolution of the battery $B_{i}$. Also indicated
is the assumed order of energy arrival and departure events from the
battery (i.e., $Y_{i}$ and $Z_{i}$) and the overflow and underflow
events (where not specified, we have $U_{i}=0$ and $O_{i}=0$).}

\label{fig:fig4}
\end{figure}

When the receiver harvests a unit of energy, $Y_{i}=1$, no energy
is used, $Z_{i}=0$, and the battery is full, $B_{i}=B_{max}$, we
have an overflow event. To keep track of the overflow events, we define
a random process $O_{i}$ such that $O_{i}=1$ if the event $\left\{ B_{i}=B_{max},Y_{i}=1,\textrm{ and }Z_{i}=0\right\} $
occurs and $O_{i}=0$ otherwise. This can be expressed as
\begin{equation}
O_{i}=1\left\{ B_{i}=B_{max},Y_{i}=1\textrm{ and }Z_{i}=0\right\} .\label{eq:O_indicator}
\end{equation}
When the receiver wishes to use a unit of energy, $Z_{i}=1$, and
both the supercapacitor and the battery are empty, $Y_{i}=0\textrm{ and }B_{i}=0$,
we have an underflow event. To describe underflow events, we introduce
a random process $U_{i}$ such that $U_{i}=1$ if the event $\left\{ B_{i}=0,Y_{i}=0\textrm{ and }Z_{i}=1\right\} $
takes place and $U_{i}=0$ otherwise. This can be expressed as 
\begin{equation}
U_{i}=1\left\{ B_{i}=0,Y_{i}=0\textrm{ and }Z_{i}=1\right\} .\label{eq:U_Indicator}
\end{equation}
A sample path of the battery state process $B_{i}$ along with $Y_{i}$,
$Z_{i}$, $U_{i}$ and $O_{i}$, is shown in Fig. \ref{fig:fig4}. 

We define the probability of underflow as 
\begin{equation}
\Pr\left\{ \mathcal{U}\right\} =\underset{n\rightarrow\infty}{\limsup}\frac{1}{n}\overset{n}{\underset{i=1}{\sum}}\textrm{E}\left[U_{i}\right],\label{eq:underflow}
\end{equation}
and the probability of overflow as 
\begin{equation}
\Pr\left\{ \mathcal{O}\right\} =\underset{n\rightarrow\infty}{\limsup}\frac{1}{n}\overset{n}{\underset{i=1}{\sum}}\textrm{E}\left[O_{i}\right].\label{eq:Overflow}
\end{equation}
We note that in (\ref{eq:underflow}) and (\ref{eq:Overflow}), the
expectation is taken over the distribution of the message $M$, of
the channel and of the receiver's energy utilization process $Z^{n}$. 

\subsection{Performance Criteria and Problem Formulation}

The point-to-point link under study will be evaluated in terms of
its performance for both information and energy transfer. A triple
$(R,P_{of},P_{uf})$ of information-energy requirements is said to
be achievable by an encoder-decoder pair if the information transfer
at rate $R$ is reliable, i.e., if 
\begin{equation}
\underset{n\rightarrow\infty}{\limsup}\,\Pr\left[\hat{M}\neq M\right]=0
\end{equation}
and if the energy transfer fulfill the constraints 
\begin{eqnarray}
\Pr\left\{ \mathcal{O}\right\}  & \leq & P_{of},\\
\textrm{and }\Pr\left\{ \mathcal{U}\right\}  & \leq & P_{uf}.
\end{eqnarray}
We are interested in investigating the set of achievable triples $(R,P_{of},P_{uf})$
for different classes of codes, namely unconstrained and $(d,k)$-RLL
constrained. To obtain further insight, in Sec. \ref{sec:Numerical-Results},
we will consider the problem 
\begin{eqnarray}
\textrm{minimize} & \max(P_{of},P_{uf})\nonumber \\
\textrm{subject to} & (R,P_{of},P_{uf}) & \textrm{is achievable},\label{eq:Opt_pr-1}
\end{eqnarray}
where $R$ is fixed and the minimization is done over all codes belonging
to a certain class. Problem (\ref{eq:Opt_pr-1}) is appropriate when
both underflow and overflow are equally undesirable and one wishes
to reduce both equally as much as possible. Alternatively, one could,
e.g., minimize either one of $P_{of}$ or $P_{uf}$ under a given
constraint on the other and on the rate.

\section{Unconstrained Codes\label{sub:Unconstrained-Codes}}

In this section, we study the information-energy transfer performance
of classical unconstrained codes. To this end, we adopt Shannon's
classical random coding argument. Accordingly, we assume that the
codewords $x^{n}(m)$, $m\in\left[1:2^{nR}\right]$, are generated
independently as i.i.d. $\verb"Ber"(p_{x})$ processes and evaluate
the corresponding performance on average over the code ensemble. As
it is well known (see, e.g., \cite{Cover}), the maximum information
rate $R$ achieved by this code is given as
\begin{eqnarray}
R & = & \textrm{I(}X;Y)\nonumber \\
 & = & \textrm{H}(Y)-\textrm{H}(Y|X)\nonumber \\
 & = & \textrm{H}(p_{x}(1-p_{10}))-p_{x}\textrm{H}(p_{10})\nonumber \\
 & = & \textrm{H}(p_{y})-\frac{p_{y}}{1-p_{10}}\textrm{H}(p_{10}),\label{eq:rate}
\end{eqnarray}
where we have defined the probability $p_{y}\triangleq\Pr[Y_{i}=1]=p_{x}(1-p_{10})$
and the binary entropy function
\begin{equation}
\textrm{H(}a)\triangleq-a\log_{2}a-(1-a)\log_{2}(1-a).\label{eq:b_entropy_function}
\end{equation}

\begin{figure}[h!]
\centering \includegraphics[bb=7bp 336bp 768bp 543bp,clip,scale=0.5]{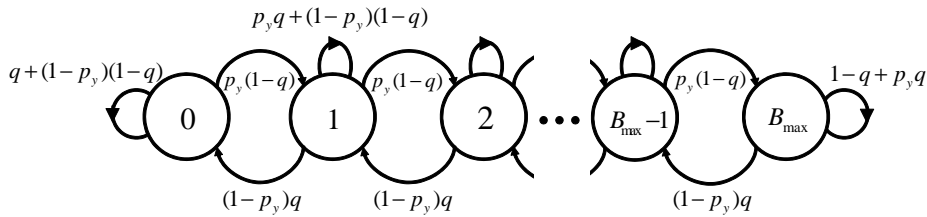}

\caption{The birth-death Markov process defining the battery state evolution
along the channel uses with unconstrained (i.i.d.) random codes and
i.i.d. receiver's energy usage process $Z^{n}$ (i.e., $q=q_{1}=1-q_{0}$).}

\label{fig:fig5}
\end{figure}

We now turn to the evolution of the performance in terms of energy
transfer. In order to simplify the analysis and obtain some insight,
we first assume the special case for then receiver's energy utilization
model in which the process $Z^{n}$ is i.i.d. and hence $q_{1}=1-q_{0}\triangleq q$.
Note that $q$ is the energy usage probability, in that we have $q=\Pr[Z_{i}=1]$.
The extension to the more general Markov model of Fig. \ref{fig:fig2}
will be discussed in Remark \ref{thm:UC_MEU}. If the process $Z^{n}$
is i.i.d., the battery state evolves according to the birth-death
Markov chain shown in Fig. \ref{fig:fig5}. Using standard considerations
and recalling (\ref{eq:O_indicator}) and (\ref{eq:U_Indicator}),
we can then calculate the probability of overflow and underflow respectively,
as
\begin{eqnarray}
\Pr\left\{ \mathcal{O}\right\}  & = & \pi_{B_{\max}}p_{y}(1-q)\triangleq\textrm{O}(p_{y}),\label{eq:overflow-1}\\
\textrm{and }\Pr\left\{ \mathcal{U}\right\}  & = & \pi_{0}(1-p_{y})q\triangleq\textrm{U}(p_{y}),\label{eq:underflow-1}
\end{eqnarray}
where $\pi_{i}$ is the steady-state probability of state $i\in\left[0,B_{\max}\right]$
for the Markov chain in Fig. \ref{fig:fig5}. This can be easily calculated
as
\begin{eqnarray}
\pi_{i} & = & \frac{A^{i}}{1+A+...+A^{B_{\max}}},
\end{eqnarray}
where $A=\frac{p_{y}(1-q)}{(1-p_{y})q}$. The following lemma summarizes
our conclusions so far.

\begin{figure}[h!]
\centering \includegraphics[clip,scale=0.5]{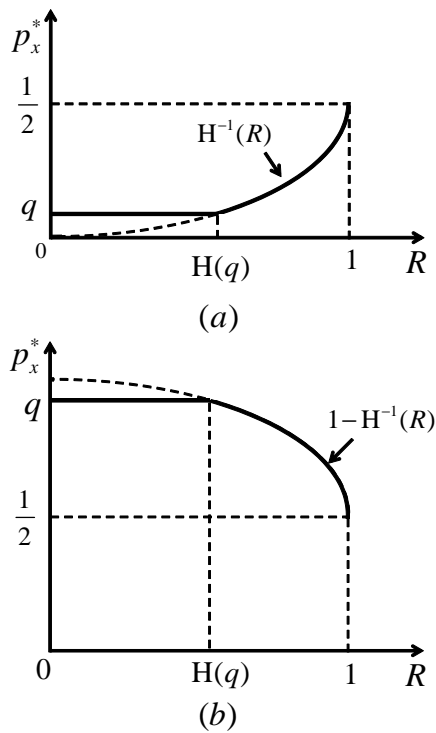}

\caption{Illustration of the optimal solution $p_{x}^{\star}$ of problem (\ref{eq:Opt_pr-2})
(Lemma \ref{lem:lemma3}): (a) the overflow-limited regime $q\leq1/2$;
(b) the underflow-limited regime $q>1/2$.}

\label{fig:fig6}
\end{figure}

\begin{lem}
\label{lem:lemma1}Given a receiver energy usage i.i.d. process with
energy usage probability $q$, the set of achievable information-energy
triples $(R,P_{of},P_{uf})$ for unconstrained (i.i.d.) codes is given
by
\begin{eqnarray}
\Biggl\{(R,P_{of},P_{uf}): & \exists & p_{y}\in\left[0,1-p_{10}\right]\textrm{ such that }\nonumber \\
 &  & R\leq\textrm{H}(p_{y})-\frac{p_{y}}{1-p_{10}}\textrm{H}(p_{10})\label{eq:Rate_region}\\
 &  & P_{of}\geq\ce{O}(p_{y}),P_{uf}\geq\ce{U}(p_{y})\Biggr\},
\end{eqnarray}
where $\ce{H}(p_{y})$, $\ce{O}(p_{y})$ and $\ce{U}(p_{y})$ are
defined in (\ref{eq:b_entropy_function}), (\ref{eq:overflow-1})
and (\ref{eq:underflow-1}), respectively. \end{lem}
\begin{rem}
The region (\ref{eq:Rate_region}) is in general not convex, but it
can be convexified if one allows for time sharing between codes with
different values of $p_{y}$ (see, e.g., \cite[Ch. 4]{El Gamal} for
related discussion). 
\end{rem}
In order to get further insight into the performance of unconstrained
codes, we now assume that the channel is noiseless, i.e., $p_{10}=0$
and, as a result, we have $Y_{i}=X_{i}$ for all $i=1,...,n$ and
$p_{y}=p_{x}$. Moreover, we consider problem (\ref{eq:Opt_pr-1}),
which reduces to the following optimization problem 
\begin{eqnarray}
\underset{p_{x}\in\left[0,1\right]}{\textrm{minimize}} & \max\left(\ce{O}(p_{x}),\ce{U}(p_{x})\right)\nonumber \\
\textrm{subject to}: & \ce{H}(p_{x})\geq R,\label{eq:Opt_pr-2}
\end{eqnarray}
where $R$ is fixed. The solution of problem (\ref{eq:Opt_pr-2})
is summarized in the following lemma.
\begin{lem}
\label{lem:lemma3}The optimal solution $p_{x}^{\star}$ of problem
(\ref{eq:Opt_pr-2}) is given as  \begin{subequations}\label{eqn: Opt_value-1}
\begin{eqnarray}
q &  & \textrm{if }R\leq\ce{H}(q)\label{eq:OPT_value1-1}\\
\ce{H^{-1}}(R) &  & \textrm{if }R>\ce{H}(q)\textrm{ and }q\leq\frac{1}{2}\label{eq:OPT_value2-1}\\
1-\ce{H^{-1}}(R) &  & \textrm{if }R>\ce{H}(q)\textrm{ and }q>\frac{1}{2}\label{eq:OPT_value3-1}
\end{eqnarray}
\end{subequations}where $\ce{H^{-1}}(R)$ is the inverse of the entropy
function in the interval $\left[0,1/2\right]$. Moreover, the optimal
value $\max\left(\ce{O}\left(p_{x}^{\star}\right),\ce{U}\left(p_{x}^{\star}\right)\right)$
of the problem (\ref{eq:Opt_pr-2}) is given by \begin{subequations}\label{eqn: Opt_value}
\begin{eqnarray}
\frac{\left(1-q\right)q}{B_{\max}+1} & \textrm{if} & R\leq\ce{H}(q)\label{eq:OPT_value1}\\
\ce{O}\left(\ce{H^{-1}}(R)\right) & \textrm{if} & R>\ce{H}(q)\textrm{ and }q\leq\frac{1}{2}\label{eq:OPT_value2}\\
\ce{U}\left(1-\ce{H^{-1}}(R)\right) & \textrm{if} & R>\ce{H}(q)\textrm{ and }q>\frac{1}{2}.\label{eq:OPT_value3}
\end{eqnarray}
\end{subequations}\end{lem}
\begin{IEEEproof}
A graphical illustration of Lemma \ref{lem:lemma3} is shown in Fig.
\ref{fig:fig6}. To interpret the conditions (\ref{eq:OPT_value1-1})
and (\ref{eq:OPT_value1}), we observe that the underflow probability
$\textrm{U}(p_{x})$ is monotonically decreasing with $p_{x}$, while
the overflow probability $\textrm{O}(p_{x})$ is monotonically increasing
with $p_{x}$. Therefore, in the absence of the rate constraint, the
optimal value of problem (\ref{eq:Opt_pr-2}) is achieved when $p_{x}=q$,
since, with this choice, we have $\ce{O}(p_{x})=\ce{U}(p_{x})$. As
a result, if $R\leq\textrm{H}(q)$, and hence the rate constraint
is immaterial for $p_{x}=q$, we have $p_{x}^{\star}=q$.

Instead, if $R>\textrm{H}(q)$, the rate constraint is active and
the optimal solution requires $R=\ce{H}(p_{x})$. In particular, there
are two situations to be considered, namely the \textit{overflow-limited
regime, }defined by the condition $q\leq1/2$, and the \textit{underflow-limited
regime,} where we have $q>1/2$. In the former regime (Fig. \ref{fig:fig6}-(a)),
the rate constraint forces $p_{x}$ to be larger than $q$, which
leads to the optimal solution $p_{x}^{\star}=\ce{H^{-1}}(R)$ and
causes the overflow probability $\textrm{O}(p_{x})$ to be larger
than the underflow probability $\textrm{U}(p_{x})$, so that $\max\left(\textrm{O}\left(p_{x}\right),\textrm{U}\left(p_{x}\right)\right)=\textrm{O}\left(p_{x}\right)$.
In contrast, in the underflow-limited regime (Fig. \ref{fig:fig6}-(b)),
the rate constraint forces $p_{x}$ to be smaller than $q$, which
leads to $p_{x}^{\star}=1-\ce{H^{-1}}(R)$ and causes $\textrm{U}\left(p_{x}\right)$
to dominate $\textrm{O}(p_{x})$, or $\max\left(\textrm{O}\left(p_{x}\right),\textrm{U}\left(p_{x}\right)\right)=\textrm{U}\left(p_{x}\right)$.\end{IEEEproof}
\begin{rem}
\label{thm:matching}The proof of Lemma \ref{lem:lemma3} suggests
that, when the rate is sufficiently small, problem (\ref{eq:Opt_pr-1})
is solved by \textquotedbl{}matching\textquotedbl{} the code structure
to the receiver's energy utilization model. This is done, under the
given i.i.d. assumption on codes and receiver's energy utilization,
by setting $p_{x}^{\star}=q$. Instead, when the rate constraint is
the limiting factor, one is forced to allow for a mismatch between
code properties and receiver's energy utilization model (by setting
$p_{x}^{\star}\neq q$). These ideas will be useful when interpreting
the gains achievable by constrained codes discussed in Sec. \ref{sub:Constrained-Codes}.
\end{rem}

\begin{figure}[h!]
\centering \includegraphics[clip,scale=0.4]{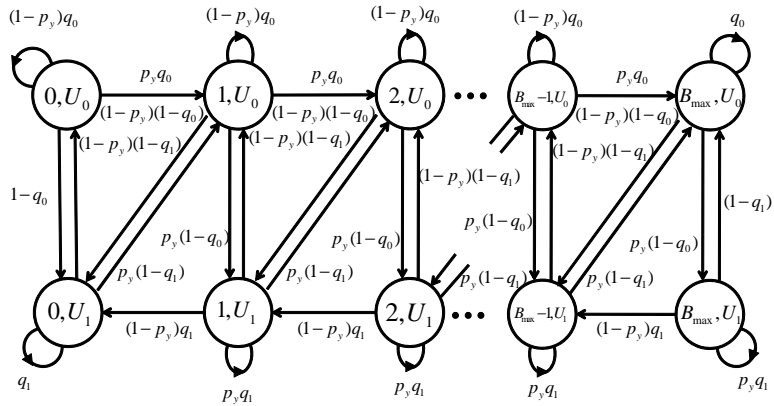}

\caption{The Markov process defining the battery state evolution along the
channel uses with unconstrained (i.i.d.) random codes and the Markov
receiver's energy usage model of Fig. \ref{fig:fig2}.}

\label{fig:fig7}
\end{figure}

\begin{rem}
\label{thm:UC_MEU}The characterization of the achievable information-energy
triples $(R,P_{of},P_{uf})$ of Lemma \ref{lem:lemma1} can be extended
to the more general Markov model in Fig. \ref{fig:fig2} for the receiver's
energy usage. This is done by noting that the battery evolution under
this model is described by the Markov chain shown in Fig. \ref{fig:fig7},
instead of the simpler birth-death Markov process shown in Fig. \ref{fig:fig5}.
The calculation of the corresponding steady-state probabilities $\pi_{i,U_{j}}$,
for $i\in\left[0,B_{\max}\right]$ and $j=\left\{ 0,1\right\} $,
can be done using standard steps (see, e.g., \cite{Gallager}). Lemma
\ref{lem:lemma1} then extends to the scenario at hand by calculating
the probabilities of overflow and underflow, similar to (\ref{eq:overflow-1})-(\ref{eq:underflow-1})
as
\begin{eqnarray}
\Pr\left\{ \mathcal{O}\right\}  & = & \pi_{B_{\max},U_{0}}p_{y}q_{0}+\pi_{B_{\max},U_{1}}p_{y}(1-q_{1})\triangleq\textrm{O}(p_{y}),\label{eq:overflow-2}\\
\textrm{and }\Pr\left\{ \mathcal{U}\right\}  & = & \pi_{0,U_{0}}(1-p_{y})(1-q_{0})+\pi_{0,U_{1}}(1-p_{y})q_{1}\triangleq\textrm{U}(p_{y}).\label{eq:underflow-2}
\end{eqnarray}

\end{rem}

\section{Constrained Codes\label{sub:Constrained-Codes}}

In this section, we study the performance of $(d,k)$-RLL codes. To
this end, as with unconstrained codes, we adopt a random coding approach.
Specifically, we take the codewords to be generated independently
according to a stationary Markov chain defined on the finite state
machine in Fig. \ref{fig:fig3}. It is known that this choice is optimal
in terms of capacity (see, e.g., \cite{Marcus,Barbero,Kofman}). A
stationary Markov chain on the graph of Fig. \ref{fig:fig3} is defined
by the transition probabilities $\mathcal{P}=\left\{ p_{d},p_{d+1},...,p_{k-1}\right\} $
on its edges as shown in Fig. \ref{fig:fig8}. We define as $C_{i}$
the state of the constrained code at time $i$, prior to the transmission
of $X_{i}$. For example, the state sequence for the type-0 $(d,k)=(2,7)$-RLL
corresponding to the codeword $x^{n}(m)=00100001001000000010$ is
$c^{n}(m)=01201234012012345670$ where $n=20$. Then, the transition
probability $p_{j}$ for $j=d,...,k-1$ is equal to $\Pr[C_{i}=j+1|C_{i-1}=j]$,
for $i>1$. Barring degenerate choices for $\mathcal{P}$, it is easy
to see that the Markov chain is irreducible, and hence one can calculate
the unique steady-state distribution $\pi_{j}=\Pr\left[C_{i}=j\right]$
for $j\in\left[0,k\right]$ (see, e.g., \cite{Marcus}).

\subsection{Information Rate}

In \cite[Lemma 5]{Zehavi}, it was proved that an achievable rate
$R$ with $(d,k)$-RLL codes is given as $R=\textrm{I}(C_{2};Y_{2}|C_{1})$.
Evaluating this expression for type-0 $(d,k)$-RLL constrained codes
leads to
\begin{eqnarray}
R & = & \textrm{H}(Y_{2}|C_{1})-\textrm{H}(Y_{2}|C_{1},C_{2})\nonumber \\
 & = & \overset{k-1}{\underset{j=d}{\sum}}\pi_{j}\left\{ \textrm{H}((1-p_{j})(1-p_{10}))-(1-p_{j})\textrm{H}(p_{10})\right\} .\label{eq:rate_noisy_constrained}
\end{eqnarray}
Instead, for type-1 code the achievable rate becomes 
\begin{eqnarray}
R & = & \overset{k-1}{\underset{j=d}{\sum}}\pi_{j}\left\{ \textrm{H}(p_{j}(1-p_{10}))-p_{j}\textrm{H}(p_{10})\right\} .\label{eq:rate_noisy_constrained-1}
\end{eqnarray}

\begin{rem}
We note that if the channel is noiseless, i.e., $p_{10}=0$, the information
rates (\ref{eq:rate_noisy_constrained}) and (\ref{eq:rate_noisy_constrained-1})
equal the entropy rate of the channel input sequence $X^{n}$, i.e.,
\begin{eqnarray}
R & = & \overset{k-1}{\underset{j=d}{\sum}}\pi_{j}\textrm{H}(p_{j}).\label{eq:rate-1-1}
\end{eqnarray}
Moreover, the maximization of the achievable information rate (\ref{eq:rate-1-1})
over the transition probabilities $\mathcal{P}$, with no regards
for energy transfer, leads to the solution 
\begin{eqnarray}
\underset{\mathcal{P}}{\sup}\overset{k-1}{\underset{j=d}{\sum}}\pi_{j}\textrm{H}(p_{j}) & = & \log_{2}\lambda,\label{eq:max_achv_rate}
\end{eqnarray}
where $\lambda$ is the largest absolute value taken by the eigenvalues
of adjency matrix%
\footnote{The adjacency matrix A is a $k\times k$ matrix such that the $(i,j)$th
element equals 1 if state $i$ and state $j$ are connected in the
graph that defines the code (see Fig. \ref{fig:fig8}) and is zero
otherwise.%
} $A$ of the graph that defines the $(d,k)$-RLL code \cite{Marcus}.
\begin{figure}[h!]
\centering \includegraphics[clip,scale=0.4]{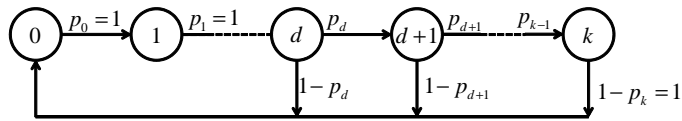}

\caption{Transition probabilities $\mathcal{P}=\left\{ p_{d},p_{d+1},...,p_{k-1}\right\} $
defining the stationary Markov chain used for random coding with type-0
$(d,k)$-RLL codes.}

\label{fig:fig8}
\end{figure}

\end{rem}

\subsection{Energy Transfer}

We now address energy transfer by turning to the calculation of the
probabilities of battery underflow and overflow, namely $\Pr\left\{ \mathcal{U}\right\} $
and $\Pr\left\{ \mathcal{O}\right\} $ in (\ref{eq:underflow}) and
(\ref{eq:Overflow}), respectively. To this end, as for unconstrained
codes, we focus at first on the special case in which the energy usage
process $Z^{n}$ is i.i.d. with energy usage probability $q$. We
refer to Remark \ref{thm:CC_MEU} below for a discussion on the extension
to the Markov model in Fig. \ref{fig:fig2}. 

\begin{figure}[h!]
\centering \includegraphics[clip,scale=0.6]{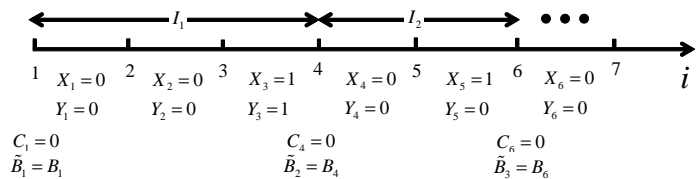} \caption{A sample of renewal events $\left\{ C_{i}=0\right\} $ and the corresponding
evolution of the battery across channel uses for a type-0 $(d,k)$-RLL
code.}

\label{fig:fig9}
\end{figure}

We use a renewal-reward argument (see, e.g., \cite{Gallager}). We
recall that a renewal process is a random process of inter-renewal
intervals $I_{1},I_{2},...$ that are positive i.i.d. random variables.
For our analysis, it is convenient to define the renewal event as
$\left\{ C_{i}=0\right\} $, so that a renewal takes place every time
the state of the constrained code $C_{i}$ is equal to 0. This is
equivalent to saying that, in the channel use before a renewal event,
the transmitted signal $X_{i}$ equals 1 for type-0 $(d,k)$-RLL codes
and $X_{i}=0$ for type-1 $(d,k)$-RLL codes. We refer to Fig. \ref{fig:fig9}
for an illustration. Based on the above, the renewal intervals $I_{i}$,
for $i\geq1$, are i.i.d. integer random variables with distribution
$p_{I}(i)$ that can be calculated, given $\mathcal{P}$, as 
\begin{equation}
p_{I}(i)=\begin{cases}
0 & i\leq d\textrm{ and }i>k+1\\
1-p_{d} & i=d+1\\
(1-p_{i-1})\underset{l=d}{\overset{i-2}{\prod}}p_{l} & d+1<i\leq k\\
\underset{l=d}{\overset{k-1}{\prod}}p_{l} & i=k+1
\end{cases}.\label{eq:PI_i}
\end{equation}
Moreover, it is useful to define a Markov chain $\tilde{B}_{i}$ that
defines the evolution of the battery as evaluated at the renewal instants
(i.e., for values of $i$ for which $C_{i}=0$), as illustrated in
Fig. \ref{fig:fig9}. We refer to the steady-state probability of
this Markov chain as $\tilde{\pi}_{b}$ with $b\in\left[0,B_{\ce{max}}\right]$.
Finally, we define as $\tilde{O}_{b}$ the random variable that counts
the number of overflow events in a renewal that starts with a battery
with capacity $b\in\left[0,B_{\ce{max}}\right]$, and, similarly,
we define as $\tilde{U}_{b}$ the random variable that counts the
number of underflow events in a renewal that starts with a battery
with capacity $b\in\left[0,B_{\ce{max}}\right]$. We proceed by treating
separately the type-0 and type-1 codes.

\subsubsection{Type-0 Codes}

For type-0 codes, the transition probabilities for the process $\tilde{B}_{i}$
are reported in Appendix \ref{sec:Apndx_A} (see also Fig. \ref{fig:fig10}
for an illustration), from which the steady state probabilities $\tilde{\pi}_{b}$
can be calculated (see, e.g., \cite{Gallager}). The next proposition
summarizes the main result of the analysis. We use the definition
$p(n;i,q)=\binom{i}{n}q^{n}(1-q)^{i-n}$ with $n=0,...,i$, for the
probability distribution of a binomial random variable with parameters
$(i,q)$.

\begin{figure}[h!]
\centering \includegraphics[clip,scale=0.5]{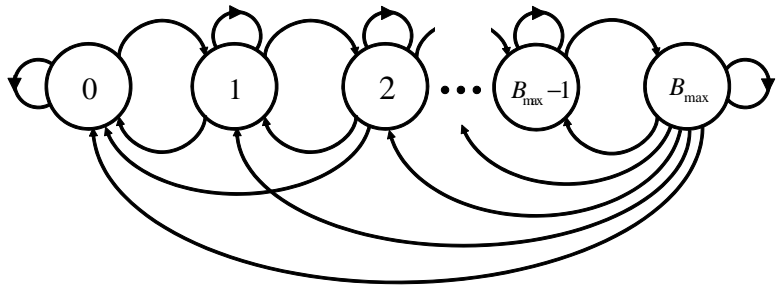}

\caption{The birth-death Markov process defining the battery state evolution
along the renewal instants, where we have $C_{i}=0$, for $(d,k)$-RLL
codes, i.i.d. receiver's energy usage pattern $Z^{n}$, and $k=B_{\max}$.}

\label{fig:fig10}
\end{figure}

\begin{prop}
\label{prop:proposition}Given an i.i.d. receiver energy usage process
with energy usage probability $q$, the set of achievable information-energy
triples $(R,P_{of},P_{uf})$ for type-0 $(d,k)$-RLL codes is given
by
\begin{eqnarray}
\Biggl\{(R,P_{of},P_{uf}): & \exists & \mathcal{P}=\left\{ p_{d},p_{d+1},...,p_{k-1}\right\} \in\left(0,1\right)^{n}\textrm{ such that }\nonumber \\
 &  & R\leq\overset{k-1}{\underset{j=d}{\sum}}\pi_{j}\left\{ \textrm{H}((1-p_{j})(1-p_{10}))-(1-p_{j})\textrm{H}(p_{10})\right\} ,\label{eq:Rate_region-1}\\
 &  & P_{of}\geq\frac{\mbox{\ensuremath{\tilde{\pi}}}_{B_{\max}}\ce{E}\left[\tilde{O}_{B_{\max}}\right]}{\ce{E}\left[I\right]},\label{eq:Overflow-1}\\
 & \textrm{and } & P_{uf}\geq\frac{\overset{B_{\max}}{\underset{b=0}{\sum}}\mbox{\ensuremath{\tilde{\pi}}}_{b}\ce{E}\left[\tilde{U}_{b}\right]}{\ce{E}\left[I\right]}\Biggr\},\label{eq:Underflow-1}
\end{eqnarray}
where we have defined\textup{
\begin{eqnarray}
\textrm{E}\left[I\right] & = & \overset{k+1}{\underset{i=d+1}{\sum}}i\cdot p_{I}(i),\label{eq:E_I}
\end{eqnarray}
}along with\textup{
\begin{eqnarray}
\textrm{E}\left[\tilde{U}_{b}\right] & = & \overset{k+1}{\underset{i=d+1}{\sum}}p_{I}(i)\left\{ (1-p_{10})\overset{i-b-1}{\underset{l=1}{\sum}}p(l+b;i-1,q)\right.\nonumber \\
 &  & \left.+p_{10}\overset{i-b}{\underset{l=1}{\sum}}p(l+b;i,q)\right\} ,\\
\textrm{and }\textrm{E}\left[\tilde{O}_{B_{\max}}\right] & = & \overset{k+1}{\underset{i=d+1}{\sum}}p_{I}(i)(1-p_{10})p(0;i,q).
\end{eqnarray}
}\end{prop}
\begin{IEEEproof}
See Appendix \ref{sec:Apndx_B}.\end{IEEEproof}
\begin{rem}
The right-hand side of (\ref{eq:Overflow-1}) evaluates the probability
of overflow as the ratio of the average numbers of overflow events
in a renewal interval over the average length of a renewal interval.
The right-hand side of (\ref{eq:Underflow-1}) can be similarly interpreted.
Note that, by the given definition of renewal events, in order to
have an overflow, the initial battery state $\tilde{B}_{i}$ must
be in state $B_{\max}$, whereas underflow events can potentially
happen for all states $b\in\{0,...,B_{\max}\}$. This is reflected
by the numerators of (\ref{eq:Overflow-1}) and (\ref{eq:Underflow-1}).
\end{rem}
\begin{figure}[h!]
\centering \includegraphics[bb=40bp 55bp 708bp 697bp,clip,scale=0.5]{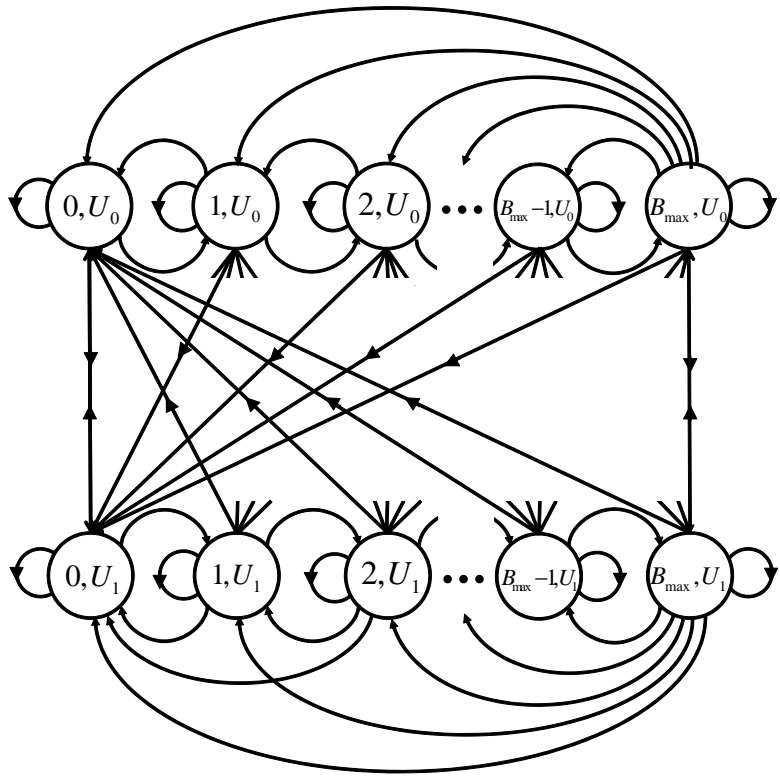}

\caption{The Markov process defining the battery state evolution along the
renewal instants where we have $C_{i}=0$, and the energy usage state
at the receiver, for $(d,k)$-RLL codes, and $k=B_{\max}$ with the
Markov receiver's energy usage model of Fig. \ref{fig:fig2}.}

\label{fig:fig11}
\end{figure}

\begin{rem}
\label{thm:CC_MEU}Similar to the case of unconstrained codes (see
Remark \ref{thm:UC_MEU}), the characterization of the achievable
information-energy triples $(R,P_{of},P_{uf})$ of Proposition \ref{prop:proposition}
can be extended to the more general Markov model in Fig. \ref{fig:fig2}
for the receiver's energy usage. This is done by noting that the evolution
of the battery state along the renewal instants can be still described
by a Markov chain, albeit a more complex one. Moreover, in order to
extend the analysis one needs to include in the state of the Markov
process not only the battery state $\tilde{B}_{i}$ but also the state
of the receiver's energy usage (either $U_{0}$ or $U_{1}$). The
corresponding Markov chain is sketched in Fig. \ref{fig:fig11}. The
calculation of the corresponding transition probabilities is straightforward
but cumbersome and is not detailed here.
\end{rem}

\subsubsection{Type-1 Codes}

For type-1 codes, the analysis presented above does not easily generalize
in the case in which the channel loss probability $p_{10}$ is nonzero.
This can be seen by following the main steps of the proof of Proposition
\ref{prop:proposition}, which is based on having at most one non-zero
received symbol per renewal interval. However, in the special case
in which $p_{10}=0$, the approach can be generalized, leading to
the following result. 
\begin{prop}
\label{prop:Proposition2}Given an i.i.d. receiver energy usage process
with energy usage probability $q$, the set of achievable information-energy
triples $(R,P_{of},P_{uf})$ for type-1 $(d,k)$-RLL codes is given
by
\begin{eqnarray}
\Biggl\{(R,P_{of},P_{uf}): & \exists & \mathcal{P}=\left\{ p_{d},p_{d+1},...,p_{k-1}\right\} \in\left(0,1\right)^{n}\textrm{ such that }\nonumber \\
 &  & R\leq\overset{k-1}{\underset{j=d}{\sum}}\pi_{j}\textrm{H}(p_{j}),\label{eq:Rate_region-1-1}\\
 &  & P_{of}\geq\frac{\overset{B_{\max}}{\underset{b=0}{\sum}}\mbox{\ensuremath{\tilde{\pi}}}_{b}\ce{E}\left[\tilde{O}_{b}\right]}{\ce{E}\left[I\right]},\label{eq:Overflow-1-1}\\
 & \textrm{and } & P_{uf}\geq\frac{\tilde{\pi}_{0}\ce{E}\left[\tilde{U}_{0}\right]}{\ce{E}\left[I\right]}\Biggr\},\label{eq:Underflow-1-1}
\end{eqnarray}
where we have defined (\ref{eq:E_I})\textup{ }along with\textup{
\begin{eqnarray}
\textrm{E}\left[\tilde{U}_{0}\right] & = & \overset{k+1}{\underset{i=d+1}{\sum}}p_{I}(i)p(0;i,1-q),\\
\textrm{and }\textrm{E}\left[\tilde{O}_{b}\right] & = & \overset{k+1}{\underset{i=d+1}{\sum}}p_{I}(i)\overset{i-b-1}{\underset{l=1}{\sum}}p(l+b;i-1,1-q).
\end{eqnarray}
}\end{prop}
\begin{IEEEproof}
Proposition \ref{prop:Proposition2} follows by the same steps as
Proposition \ref{prop:proposition} and is not detailed here.\end{IEEEproof}
\begin{rem}
Similar to Proposition \ref{prop:proposition}, the right-hand sides
of (\ref{eq:Overflow-1-1}), and of (\ref{eq:Underflow-1-1}), evaluate
the probabilities of overflow, and of underflow, via the ratios of
the average numbers of overflow, and of underflow, events in a renewal
interval over the average length of a renewal interval. In a dual
fashion with respect to Proposition \ref{prop:proposition}, given
the definition of renewal events, underflow can only occur in renewal
intervals with initial battery state $\tilde{B}_{i}$ is zero, whereas
overflow events can potentially happen for all states $b\in\{0,...,B_{\max}\}$.
\end{rem}

\section{Numerical Results\label{sec:Numerical-Results}}

\begin{figure}[h!]
\centering \includegraphics[clip,scale=0.5]{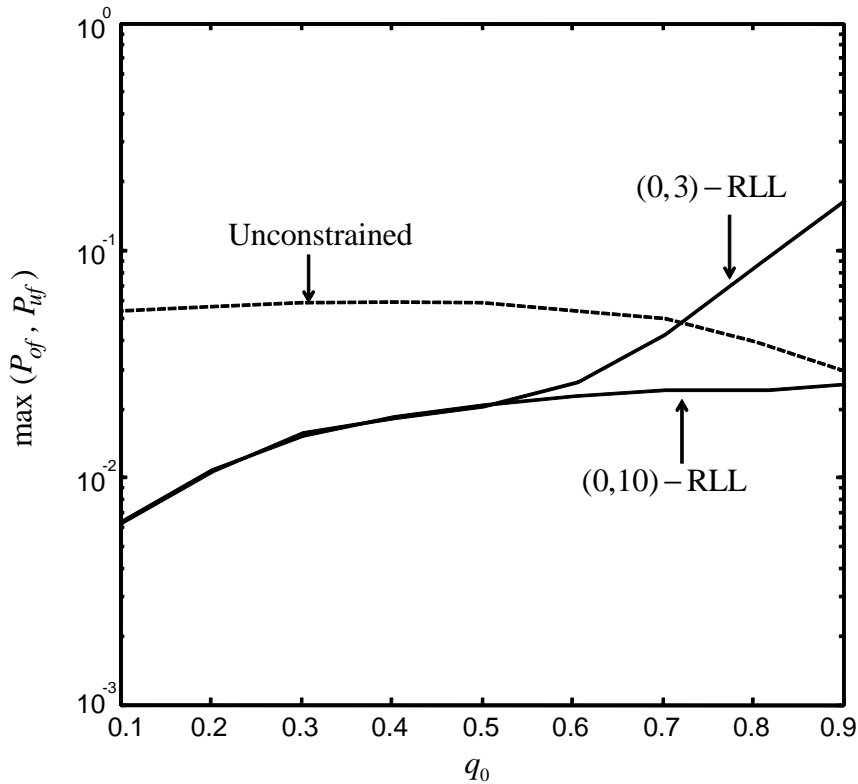}

\caption{Maximum between probability of underflow $P_{uf}$ and overflow $P_{of}$
as per problem (\ref{eq:Opt_pr-1}) for unconstrained and type-0 constrained
codes versus $q_{0}$ with $q_{1}=0$ (see Fig. \ref{fig:fig2}) and
$R=0.1$. To simplify the numerical optimization, the curve for $k=10$
has been obtained by optimizing only over $p_{0}$, $p_{1}$, $p_{2}$,
$p_{3}$ and $p_{9}$ in $\mathcal{P}=\left\{ p_{0},p_{1},...,p_{9}\right\} $
and setting $p_{3}=p_{4}=p_{5}=...=p_{8}$.}

\label{fig:fig12}
\end{figure}

In this section, we compare the performance of unconstrained and constrained
codes using problem (\ref{eq:Opt_pr-1}) as the benchmark. Fig. \ref{fig:fig12}
shows the optimal value of $\max(P_{of},P_{uf})$ for a noiseless
channel, i.e., $p_{10}=0$ in Fig. \ref{fig:fig1}, when $R=0.1$
and $q_{1}=0$ versus $q_{0}$ (recall Fig. \ref{fig:fig2}). With
$q_{1}=0$, the energy usage process $Z^{n}$ is such that a single
energy request (i.e., $Z_{i}=1$) is followed by an average of $1/(1-q_{0})$
instants where no energy is required (i.e., $Z_{i}=0$). Therefore,
as $q_{0}$ increases from 0.1 to 0.9, the average length of an interval
with no energy usage increases from around 1 to 10. Similar to the
discussion in Remark \ref{thm:matching} for unconstrained codes,
when neglecting the rate constraint, problem (\ref{eq:Opt_pr-1})
is observed to be optimized by matching the code structure to the
receiver's energy utilization model. When $q_{0}$ is sufficiently
small, this can be easily accomplished with type-0 $(d,k)$-RLL codes
with a small $k$. This is because $k$ defines the maximum possible
number of zero symbols $X_{i}$ sent before a symbol $X_{i}=1$. As
$q_{0}$ increases, and hence the average length of the bursts of
zeros grows in the process $Z_{i}$, the value of $k$ must be correspondingly
increased. This is confirmed by Fig. \ref{fig:fig12}, which shows
the significant gain achievable by the use of RLL codes when properly
selecting the code parameters. We observe that type-1 $(d,k)$-RLL
codes would provide exactly the same performance in the symmetric
case in which we have $q_{0}=0$, and hence intervals of energy usage
(i.e., $Z_{i}=1$) are followed by a single instant with no energy
usage (i.e., $Z_{i}=0$). 

\begin{figure}[h!]
\centering \includegraphics[clip,scale=0.5]{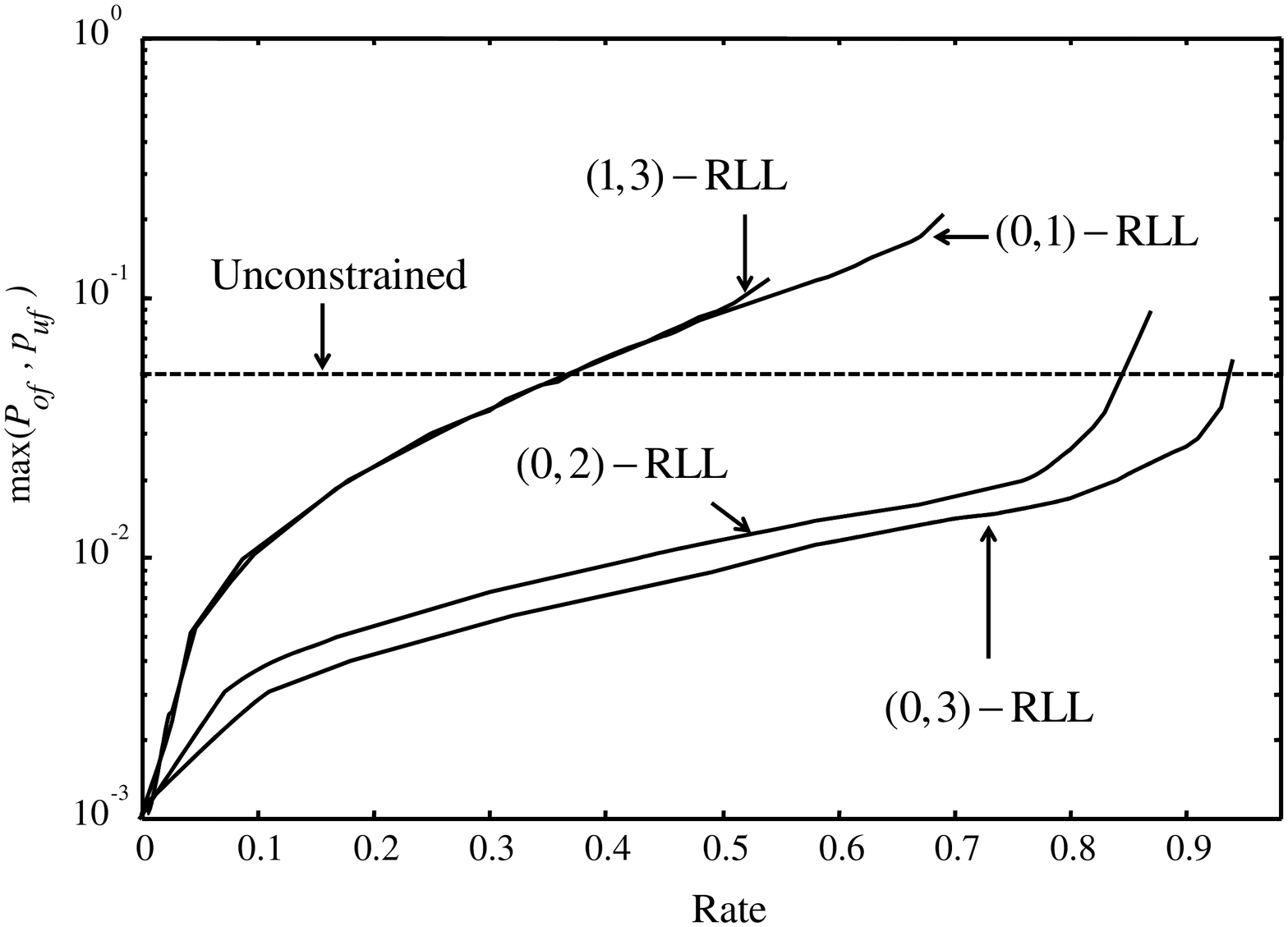}

\caption{Maximum between probability of underflow and overflow as per problem
(\ref{eq:Opt_pr-1}) for unconstrained and type-0 constrained codes
versus the information rate $R$ with $q_{0}=q_{1}=0$ (see Fig. \ref{fig:fig2}). }

\label{fig:fig13}
\end{figure}

The impact of the information rate $R$ is illustrated in Fig. \ref{fig:fig13}
for $q_{0}=q_{1}=0$ and $p_{10}=0$. Following the discussion above,
when the rate is small, with $q_{0}=q_{1}=0$, it is sufficient to
choose a type-0 or type-1 $(d,k)$-RLL code with $k=1$, as this matches
the energy usage process. However, as the rate grows larger, one needs
to increase the value of $k$, while keeping $d$ as small as possible
\cite{Marcus}. For instance, with $k=1$ and $d=0$, the maximum
achievable rate (\ref{eq:max_achv_rate}) is $R=0.6942$; with $k=2$
and $d=0$, it is $R=0.8791$; with $k=3$ and $d=1$, it is $R=0.5515$;
and with $k=3$ and $d=0$, it is $R=0.9468$ \cite[Table 3.1]{Marcus}.
Accordingly, Fig. \ref{fig:fig13} shows again that, by appropriately
choosing $d$ and $k$, RLL codes can provide relevant advantages.

\begin{figure}[h!]
\centering \includegraphics[clip,scale=0.5]{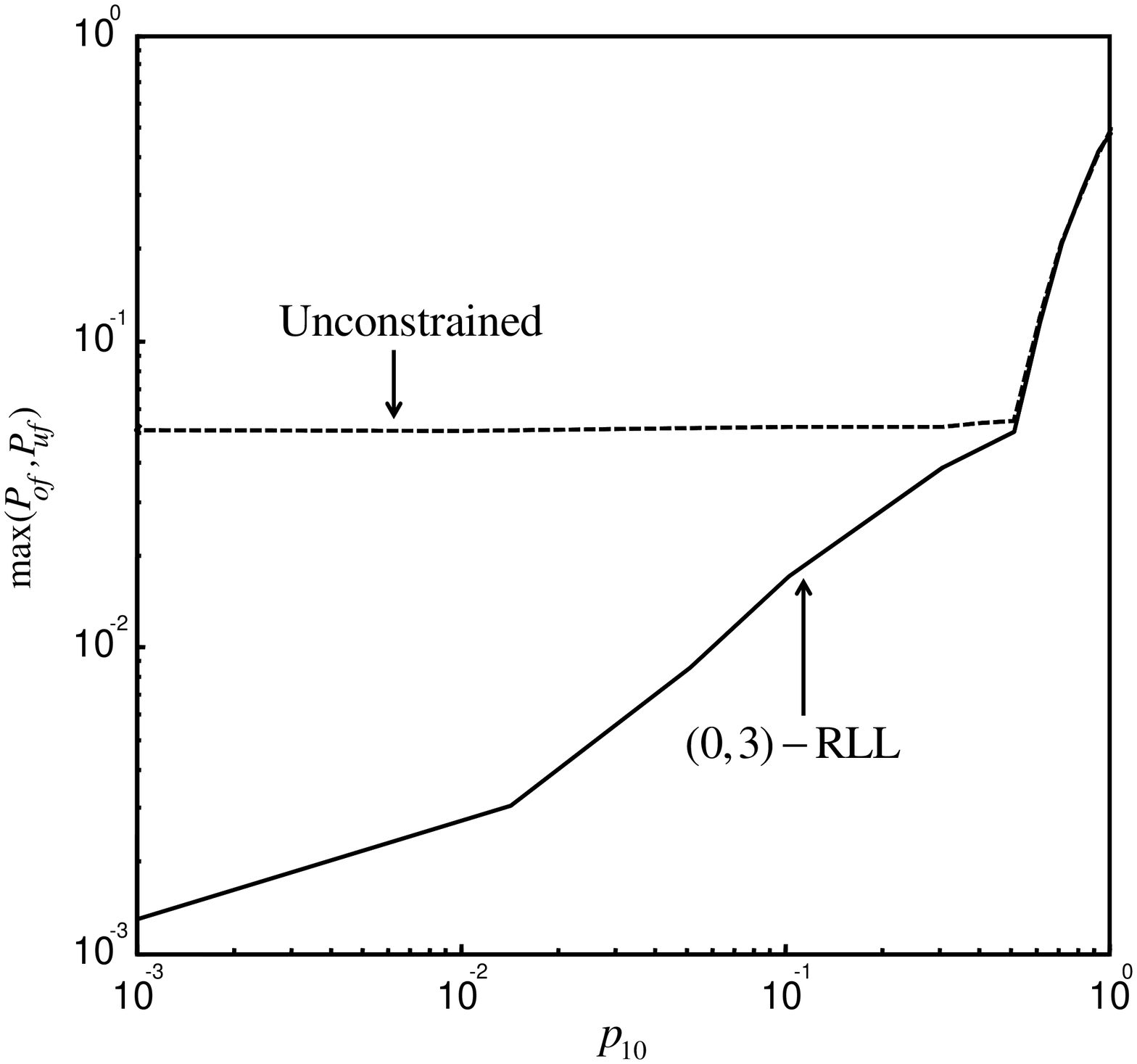}

\caption{Maximum between probability of underflow and overflow as per problem
(\ref{eq:Opt_pr-1}) for unconstrained and constrained codes versus
$p_{10}$ for $k=3$, $R=0.01$ and $q_{0}=q_{1}=0$. }

\label{fig:fig14}
\end{figure}

Finally, we observe the effect of the loss probability $p_{10}$ in
Fig. \ref{fig:fig14} where we set $R=0.01$ and $q_{0}=q_{1}=0$.
Following the discussion above (see Remark \ref{thm:matching}), in
order to match the receiver's energy utilization, the unconstrained
code should be designed in such a way that $p_{y}=0.5$ since $\Pr[Z_{i}=1]=0.5$.
Given that $p_{y}=p_{x}(1-p_{10})$, this is only possible for $p_{10}<0.5$,
and hence, for $p_{10}>0.5$, the performance is degraded as seen
in Fig. \ref{fig:fig14}. When $p_{10}=0$, as demonstrated above,
RLL codes provide significant gains by providing a better matching
to the utilization process $Z^{n}$. As the losses on the channel
become more pronounced this gain decreases due to the reduced control
of the received signal afforded by designing the transmitted signal.

\section{Conclusions\label{sec:Conclusions}}

A host of new applications, including body area networks with implantable
devices, is enabled by the possibility to reuse the energy received
from information-bearing signals. With these applications in mind,
we have investigated the use of constrained run-length limited (RLL)
codes with the aim of enhancing the achievable performance in terms
of simultaneous information and energy transfer. We have proposed
a framework whereby the performance of energy transfer is measured
by the probabilities of underflow and overflow at the receiver. The
analysis has demonstrated that constrained codes enable the transmission
strategy to be better adjusted to the receiver's energy utilization
pattern as compared to classical unstructured codes. This has been
shown to lead to significant performance gains especially at low information
rates. Interesting future work includes the investigation of non-binary
codes and multi-terminal scenarios.

\appendices{}

\section{Transition Probabilities for The Markov Chain in Fig. \ref{fig:fig10}
\label{sec:Apndx_A}}

Using the definitions in Sec. \ref{sub:Constrained-Codes}, we now
calculate the transition probability $\tilde{p}_{m,m-n}\triangleq\Pr[\tilde{B}_{i}=m-n|\tilde{B}_{i-1}=m]$,
where $\tilde{B}_{i}$ is the random process that describes the evolution
of the battery at the renewal instants (see Fig. \ref{fig:fig9}).
The probabilities $\tilde{p}_{m,m-n}$ for $m\in\left[0,B_{\ce{max}}\right]$
and $n\in\left[-1,m\right]$ can be calculated as 
\begin{eqnarray}
\underset{m\neq0,1,B_{\max}}{\tilde{p}_{m,m-n}}\negmedspace\negmedspace & \negmedspace\negmedspace=\negmedspace\negmedspace & \negmedspace\negmedspace\negmedspace\negmedspace\overset{k+1}{\underset{i=d+1}{\sum}}\negmedspace\negmedspace p_{I}(i)\negmedspace\cdot\negmedspace\begin{cases}
\negmedspace\negmedspace(1\negmedspace-\negmedspace p_{10})p(0;i,q)\negmedspace\negmedspace & n=-1\negmedspace\negmedspace\\
\negmedspace\negmedspace(1\negmedspace-\negmedspace p_{10})q\overset{i-1}{\underset{l=n}{\sum}}p(l;i\negmedspace-\negmedspace1,q)+p_{10}\overset{i}{\underset{l=n}{\sum}}p(l;i,q)\negmedspace\negmedspace & n=m\negmedspace\negmedspace\\
\negmedspace\negmedspace(1\negmedspace-\negmedspace p_{10})\left[qp(n;i-1,q)\negmedspace\negmedspace+\negmedspace\negmedspace(1\negmedspace-\negmedspace q)p(n+1;i-1,q)\right] & \negmedspace\negmedspace\\
\negmedspace\negmedspace+p_{10}p(n;i,q)\negmedspace\negmedspace & n=0,1,...,m\negmedspace-\negmedspace1\negmedspace\negmedspace
\end{cases}\negmedspace\negmedspace\negmedspace\negmedspace,
\end{eqnarray}

\begin{eqnarray}
\underset{m=0}{\tilde{p}_{m,m-n}}\negmedspace\negmedspace & \negmedspace\negmedspace=\negmedspace\negmedspace & \negmedspace\negmedspace\overset{k+1}{\underset{i=d+1}{\sum}}\negmedspace p_{I}(i)\negmedspace\cdot\negmedspace\begin{cases}
(1-p_{10})(1-q)\overset{i-1}{\underset{l=n+1}{\sum}}p(l;i-1,q) & n=-1\\
(1-p_{10})q\overset{i-1}{\underset{l=n}{\sum}}p(l;i-1,q)\negmedspace+\negmedspace p_{10}\overset{i}{\underset{l=n}{\sum}}p(l;i,q) & n=0
\end{cases}\negmedspace\negmedspace,\nonumber \\
 & = & \overset{k+1}{\underset{i=d+1}{\sum}}p_{I}(i)\cdot\begin{cases}
(1-p_{10})(1-q) & n=-1\\
(1-p_{10})q+p_{10} & n=0
\end{cases},
\end{eqnarray}

\begin{eqnarray}
\underset{m=1}{\tilde{p}_{m,m-n}}\negmedspace\negmedspace & \negmedspace=\negmedspace & \negmedspace\negmedspace\overset{k+1}{\underset{i=d+1}{\sum}}\negmedspace\negmedspace p_{I}(i)\negmedspace\cdot\negmedspace\begin{cases}
\negmedspace\negmedspace(1\negmedspace-\negmedspace p_{10})p(0;i,q) & n=-1\\
\negmedspace\negmedspace(1\negmedspace-\negmedspace p_{10})\left[qp(n;i-1,q)+(1\negmedspace-\negmedspace q)\overset{i-1}{\underset{l=n+1}{\sum}}p(l;i\negmedspace-\negmedspace1,q)\right]\\
\negmedspace\negmedspace+p_{10}p(0;i,q) & n=0\\
\negmedspace\negmedspace(1\negmedspace-\negmedspace p_{10})q\overset{i-1}{\underset{l=n}{\sum}}p(l;i\negmedspace-\negmedspace1,q)+p_{10}\overset{i}{\underset{l=n}{\sum}}p(l;i,q) & n=1
\end{cases}\negmedspace\negmedspace,
\end{eqnarray}
and

\begin{eqnarray}
\underset{m=B_{\max}}{\tilde{p}_{m,m-n}}\negmedspace\negmedspace & \negmedspace\negmedspace\negmedspace=\negmedspace\negmedspace\negmedspace & \negmedspace\negmedspace\negmedspace\negmedspace\overset{k+1}{\underset{i=d+1}{\sum}}\negmedspace\negmedspace p_{I}(i)\negmedspace\cdot\negmedspace\begin{cases}
\negmedspace\negmedspace(1\negmedspace-\negmedspace p_{10})\left[p(0;i,q)+qp(0;i\negmedspace-\negmedspace1,q)\negmedspace\negmedspace+\negmedspace\negmedspace(1\negmedspace-\negmedspace q)p(1;i\negmedspace-\negmedspace1,q)\right]\negmedspace\negmedspace\negmedspace\negmedspace & \negmedspace\negmedspace\negmedspace\negmedspace\\
\negmedspace\negmedspace+p_{10}p(0;i,q)\negmedspace\negmedspace\negmedspace\negmedspace & \negmedspace n=0\negmedspace\negmedspace\negmedspace\negmedspace\\
\negmedspace\negmedspace(1\negmedspace-\negmedspace p_{10})\left[qp(n;i\negmedspace-\negmedspace1,q)\negmedspace\negmedspace+\negmedspace\negmedspace(1-q)p(n\negmedspace+\negmedspace1;i\negmedspace-\negmedspace1,q)\right]\negmedspace\negmedspace\negmedspace\negmedspace & \negmedspace\negmedspace\negmedspace\negmedspace\\
\negmedspace\negmedspace+p_{10}p(n;i,q)\negmedspace\negmedspace\negmedspace\negmedspace\negmedspace\negmedspace & \negmedspace n\negmedspace=1,2,...m\negmedspace-\negmedspace2\negmedspace\negmedspace\\
\negmedspace\negmedspace(1\negmedspace-\negmedspace p_{10})\left[qp(n;i\negmedspace-\negmedspace1,q)\negmedspace\negmedspace+\negmedspace\negmedspace(1\negmedspace-\negmedspace q)\overset{i-1}{\underset{l=n+1}{\sum}}p(l;i\negmedspace-\negmedspace1,q)\right]\negmedspace\negmedspace\negmedspace\negmedspace & \negmedspace\negmedspace\negmedspace\negmedspace\negmedspace\\
\negmedspace\negmedspace+p_{10}p(n;i,q)\negmedspace\negmedspace\negmedspace\negmedspace & \negmedspace n=m\negmedspace-\negmedspace1\negmedspace\negmedspace\negmedspace\negmedspace\\
\negmedspace\negmedspace(1\negmedspace-\negmedspace p_{10})q\overset{i-1}{\underset{l=n}{\sum}}p(n;i\negmedspace-\negmedspace1,q)\negmedspace\negmedspace+\negmedspace\negmedspace p_{10}\overset{i}{\underset{l=n}{\sum}}p(n;i,q)\negmedspace\negmedspace\negmedspace\negmedspace & \negmedspace n=m\negmedspace\negmedspace\negmedspace\negmedspace
\end{cases}.
\end{eqnarray}

\section{Proof of Proposition \ref{prop:proposition}\label{sec:Apndx_B}}

In this Appendix, we prove Proposition \ref{prop:proposition} following
a similar approach as \cite[Theorem. 5.4.1]{Gallager}. We first relate
the overflow event (\ref{eq:O_indicator}) and the underflow event
(\ref{eq:U_Indicator}) to the processes $\tilde{O}_{b}$ and $\tilde{U}_{b}$
that count the number of overflow and underflow events across the
renewal intervals (recall Sec. \ref{sub:Constrained-Codes}). To this
end, we define a random process that counts the number of renewals
(i.e., events $\{C_{j}=0\}$) up to time $i$, namely
\begin{equation}
N(i)=\left|\left\{ j\in\{1,...,i\}:C_{j}=0\right\} \right|,\label{eq:N_i}
\end{equation}
where $\left|\cdot\right|$ represents the cardinality of its argument.
It is also convenient to classify the renewal events depending on
the value of the battery at the beginning of the renewal interval.
We can then define

\begin{equation}
N_{b}(i)=\left|\left\{ j\in\{1,...,i\}:C_{j}=0\textrm{ and }B_{j}=b\right\} \right|.\label{eq:N_b_i}
\end{equation}
The relationship between (\ref{eq:N_i}) and (\ref{eq:N_b_i}) is
given as
\begin{equation}
N(i)=\underset{b=0}{\overset{B_{\max}}{\sum}}N_{b}(i).
\end{equation}
Moreover, the initial time instant of the $i$th interval corresponding
to an initial battery state $b\in\{0,...,B_{\max}\}$ can be written
as
\begin{equation}
S_{b,j}=\min\left\{ i:N_{b}(i)=j\right\} .\label{eq:S_b_j}
\end{equation}
Using (\ref{eq:N_i})-(\ref{eq:S_b_j}), we can now obtain the relationship
(see also \cite[pp. 239-240]{Gallager})
\begin{eqnarray}
\frac{\overset{B_{\max}}{\underset{b=0}{\sum}}\overset{N_{b}(i)}{\underset{j=1}{\sum}}\tilde{U}_{b,j}}{i}\leq & \frac{\overset{i}{\underset{j=1}{\sum}}U_{j}}{i} & \leq\frac{\overset{B_{\max}}{\underset{b=0}{\sum}}\overset{N_{b}(i)+1}{\underset{j=1}{\sum}}\tilde{U}_{b,j}}{i},\label{eq:inequality}
\end{eqnarray}
where 
\begin{eqnarray}
\tilde{U}_{b,j} & = & \overset{S_{b,j}}{\underset{k=S_{b,j-1}}{\sum}}U_{k}.
\end{eqnarray}
Averaging over all battery states $b$, we also have 
\begin{equation}
\textrm{E}\left[\tilde{U}_{b}\right]=\overset{k+1}{\underset{i=d+1}{\sum}}p_{I}(i)\textrm{E}\left[\tilde{U}_{b,i}\right].
\end{equation}
The left hand side of (\ref{eq:inequality}) can be separated as
\begin{eqnarray}
\frac{\overset{B_{\max}}{\underset{b=0}{\sum}}\overset{N_{b}(i)}{\underset{j=1}{\sum}}\tilde{U}_{b,j}}{i}= & \frac{\overset{B_{\max}}{\underset{b=0}{\sum}}\overset{N_{b}(i)}{\underset{j=1}{\sum}}\tilde{U}_{b,j}}{N(i)} & \frac{N(i)}{i}.\label{eq:separation}
\end{eqnarray}
Therefore, $t\rightarrow\infty$, since we have $N(t)\rightarrow\infty$,
the strong law of renewal processes can be invoked on the second term
on the right hand side of (\ref{eq:separation}) to conclude that
$N(i)/i\rightarrow1/\textrm{E}\left[I\right]$ with probability one
\cite{Gallager}. As for the first term, it can be written as
\begin{eqnarray}
\frac{\overset{B_{\max}}{\underset{b=0}{\sum}}\overset{N_{b}(i)}{\underset{j=1}{\sum}}\tilde{U}_{b,j}}{N(i)} & = & \overset{B_{\max}}{\underset{b=0}{\sum}}\frac{\overset{N_{b}(i)}{\underset{j=1}{\sum}}\tilde{U}_{b,j}}{\overset{B_{\max}}{\underset{b^{\prime}=0}{\sum}}N_{b^{\prime}}(i)}\\
 & = & \overset{B_{\max}}{\underset{b=0}{\sum}}\left(\frac{\overset{N_{b}(i)}{\underset{j=1}{\sum}}\tilde{U}_{b,j}}{N_{b}(i)}\frac{N_{b}(i)}{\overset{B_{\max}}{\underset{b^{\prime}=0}{\sum}}N_{b^{\prime}}(i)}\right)\label{eq:20}
\end{eqnarray}
As a result, if $t\rightarrow\infty$, and hence $N_{b}(i)\rightarrow\infty$,
by the strong law of large numbers, noting the fact that the random
variables $\tilde{U}_{b,j}$ for every $b\in\left[0,B_{\max}\right]$
are i.i.d. across $j$, we have $\sum_{j=1}^{N_{b}(i)}\tilde{U}_{b,j}/N_{b}(i)$$\rightarrow\textrm{E}[\tilde{U}_{b,j}]$
with probability one. Finally, by the law of large numbers for ergodic
Markov chains (see, e.g., \cite{Marcus}), we have $N_{b}(i)/\sum_{b^{\prime}=0}^{B_{\max}}N_{b^{\prime}}(i)$$\rightarrow\tilde{\pi}_{b}$,
where we recall that $\tilde{\pi}_{b}$ is the steady-state of the
Markov chain $\tilde{B}_{i}$, which can be calculated from the transition
probabilities detailed in Appendix \ref{sec:Apndx_A}. From the discussion
above, we conclude that the following limit holds with probability
one 
\begin{eqnarray}
\underset{i\rightarrow\infty}{\lim}\frac{\overset{B_{\max}}{\underset{b=0}{\sum}}\overset{N_{b}(i)}{\underset{j=1}{\sum}}\tilde{U}_{b,j}}{i} & =\frac{\overset{B_{\max}}{\underset{b=0}{\sum}}\left(\textrm{E}\left[\tilde{U}_{b}\right].\tilde{\pi}_{b}\right)}{\textrm{E}\left[I\right]} & .
\end{eqnarray}
The same limit is obtained by applying the approach detailed above
to the right-hand side of the inequality (\ref{eq:inequality}). This
concludes the proof of (\ref{eq:Underflow-1}) in Proposition \ref{prop:proposition}.
The overflow probability (\ref{eq:Overflow-1}) is obtained following
the same approach.

\end{document}